# The Space Sustainability Paradox

*Dr. Andrew Ross Wilson*

Aerospace Centre of Excellence, Department of Mechanical & Aerospace Engineering, University of Strathclyde, James Weir Building, 75 Montrose Street, Glasgow, G1 1XJ, United Kingdom, andrew.r.wilson@strath.ac.uk

*Prof. Massimiliano Vasile*

Aerospace Centre of Excellence, Department of Mechanical & Aerospace Engineering, University of Strathclyde, James Weir Building, 75 Montrose Street, Glasgow, G1 1XJ, United Kingdom, massimiliano.vasile@strath.ac.uk

**Abstract**

The recent growth of the space sector, spurred by a surge in private actors, has led to a sharp increase in our ability to address societal challenges through space data. However, this has exacerbated an already critical situation in space: the proliferation of space debris and a critical expansion of space traffic which is leading to high levels of orbital congestion. In parallel, increased levels of spacecraft production and orbital launches are also heightening the environmental footprint of the sector. This might lead to a paradoxical situation whereby the use of space to support the Sustainable Development Goals (SDGs) becomes unsustainable from the perspective of both the Earth and space environment. This situation can be described as the 'space sustainability paradox'. This paper presents this concept for the first time and argues that existing policies and remediation actions are not a long-term sustainable solution for tackling this issue and may actually intensify the problem. This places an added importance upon addressing space sustainability in a more coherent, strategic and responsible manner, potentially based on the doughnut economic model of social and planetary boundaries. Doing so may prevent the sector from falling victim to a 'tragedy of the commons' type of scenario and avoid negative trends from becoming the norm. As a result, this would ensure that outer space can continue to be used by future generations to address societal challenges, without priming severe and enduring damage to the Earth and space environment.

## 1. Introduction

The concept of sustainability has grown to a point where it is now a key innovation driver for many businesses to remain relevant and competitive in the current economic climate. Within the space industry, this shift has been a lot slower than in other sectors, particularly due to space systems traditionally being key granted exemptions from several sustainability policies. Specific space sustainability legislation and guidance itself has also been weak or lacking on many fronts, essentially allowing the space industry to operate as normal without being held accountable for irresponsible behaviour [1]. This has led to over-populated orbits and the accumulation of space debris, which threatens the future long-term accessibility to space.

This is particularly problematic since the expansion of services related to Earth Observation (EO) and Global Navigation Satellite Systems (GNSS) which now play a critical role in supporting the attainment of the 17 SDGs. The importance of these data sources are specifically recognised by the United Nations (UN) in the 2030 Agenda for Sustainable Development [2]. In a joint study by the United Nations Office for Outer Space Affairs (UNOOSA) and the European Global Navigation Satellite System (GNSS) Agency called 'Space4SDGs', it was found all 17 SDGs are positively impacted by the benefits of space, with 65 out of 169 targets (almost 40%) benefitting from European GNSS (EGNSS) and Copernicus missions alone [3]. The use of such data is also increasingly being commercialised to benefit society as a whole in a variety of areas including infrastructure, business, communications, etc. Clearly, with such a high reliance on space data to address sustainable development issues on Earth, there is a clear case for space environment management and space sustainability.

Heightened public interest in such issues were recently stimulated by the recent spaceflights of Sir Richard Branson and Jeff Bezos as well as several other news items on future proposed mega-constellations [4, 5, 6, 7]. These events, coupled with increased press coverage of other space missions, the space debris issue and sustainability of other industries, has swiftly raised a vexing question: how sustainable actually is the space sector? Such lines of questioning have led to a growing realisation within the sector that it is no longer considered acceptable for industrial stakeholders to completely disregard sustainability issues (as was previously the case for many years).

As a response, the space sector has attempted to justify the positive role of space applications through the benefits that they provide humanity [8]. However, this justification neglects the impacts that space activities have on the space environment through the crowding of critical orbits or mounting space debris population. It also discards the impacts that space activities have on Earth which can be seen as the sum of activities over the entire life cycle of a satellite (i.e. design, production, testing, transportation, launch, use and re-entry) which collectively are a source of environmental pollution.

Ultimately, there is a need for more sustainable interventions and practices which consider the full spectrum of sustainability issues associated with space. However, without an understanding on what space sustainability is, such interventions may actually have severe unintended consequences whereby the proposed interventions actually do more damage than good. As such, this paper will introduce a new concept called 'the space sustainability paradox to describe the situation where space activities which aim to positively contribute to sustainable development may actually have a detrimental impact. To do this, a review will be conducted on the current understanding of space sustainability as a definition before going on to discuss the current state of affairs in the space sector, including proposed space sustainability measures and strategies. It will then go on to define the space sustainability paradox, presenting the current problem including the range of paradoxical issues associated with the term 'space sustainability'. Thereafter, the paper will evaluate the adequacy of current and proposed space sustainability measures and strategies, outlining important rethinks that must take place if the sector is to avoid falling victim to the space sustainability paradox.

2. **Defining Space Sustainability**

Although 'space sustainability' is becoming a frequently used expression within the space sector, until recently, there was no widely agreed definition of the term. To address this, in 2018, the United Nations (UN) Committee on the Peaceful Uses of Outer Space (COPUOS) set out guidelines for the Long-term Sustainability of Outer Space Activities [9]. These guidelines provided a preliminary definition for space sustainability for the first time, as:

*"The long-term sustainability of outer space activities is defined as the ability to maintain the conduct of space activities indefinitely into the future in a manner that realizes the objectives of equitable access to the benefits of the exploration and use of outer space for peaceful purposes, in order to meet the needs of the present generations while preserving the outer space environment for future generations."*

Despite expressing a definition for the term, the contents of the guidance mainly focus on preserving the space environment. Similarly, when space sustainability is discussed within the sector, attention usually turns to just two aspects: space data and space debris.

More specifically on these aspects, space technologies and applications have played a significant role in reducing environmental impacts, promoting social well-being and fostering economic growth on Earth. From an environmental perspective, one example is Earth Observation (EO) missions, where data is collected and transmitted from spacecraft so that sustainability issues can be addressed terrestrially on Earth [10, 11]. Therefore, the development of new space missions can have a direct contribution to the success of different environmental agreements by providing arrange of services such as the provision of additional knowledge, science-based analyses, environmental monitoring or early warnings. As a result, such missions can also be aligned with the goals contained in the 2030 Agenda which is typically the approach used by the space sector for declaring their contribution towards sustainable development [12]. The role of EO and geolocation (provided by GNSS) in supporting the achievement of the SDGs is formally recognised by the 2030 Agenda for Sustainable Development (A/RES/70/1) [2]. Recent research evidences the fact that space programmes and technologies positively contribute to all 17 SDGs and hence, play a key role in the urgent global endeavour to achieve sustainable development [13]. However, increased exploitation of space is causing these orbits to carry a distinct risk of congestion, collisions and ultimately space debris. Space debris is the accumulation of man-made objects in-orbit around Earth from over 60 years of space exploration [14]. It is a serious and dangerous problem as the growth of debris could render space inaccessible for the foreseeable future. Unless it is urgently addressed, this has the potential to effectively end industry operations [15]. The reason for this is described through the Kessler syndrome which is a mathematical singularity where the population of orbital space debris is high enough that (even without any more launches) collisions between objects could cause a cascading effect, generating more debris and significantly increasing the likelihood of further collisions [16].

Whilst the importance of the previously mentioned issues should not be understated, limiting the scope of space sustainability to refer exclusively to them is quite restrictive and absolves the space sector from responsibility for the direct impact of their operations. In this regard, viewing space activities as a possible source of pollution to Earth is a topic which is commonly ignored or considered as a mere afterthought of other space sustainability initiatives, despite a growing abundance of research in this area. In particular, space life cycle assessment (LCA) is now becoming a powerful tool for calculating the environmental impacts of space activities over their lifespan [17]. Regardless, guidance documents on space sustainability are generally devoid of such practices. For example, the Secure World Foundation "Space sustainability – a practical guide" [11] mentions nothing on terrestrial-based environmental impacts at all. As a result, there has been growing calls for space environmentalism [18, 19], which is particularly crucial given Miraux [20] suggests that the space sector is bound by environmental limits.

In response to this issue, Space Scotland created an Environmental Task Force (ETF) which was tasked to develop a sustainability and net-zero roadmap for Scottish space sector (believed to be a world's first) [21]. In their quest to develop this roadmap, the ETF decided to adopt the UN COPUOS definition of space sustainability as a starting point for defining space sustainability in a Scottish Context. However, the definition was quickly adapted due to its heavy focus on the sustainability of activities in space, which neglected the sustainability of space activities themselves. The new definition proposed by the ETF was discussed at a variety of industry and stakeholder engagement events and was amended to better align with both current practice in Scotland and with the broader aims of the sustainability and net-zero roadmap. The following definition was agreed (with all changes to the UN COPUOS definition highlighted in bold):

*"The long-term sustainability of outer space activities (**on-ground and in-orbit**) is defined as the ability to maintain **and improve** the conduct of space activities indefinitely into the future in a manner that ~~realizes the objectives of equitable~~ **ensures continued** access to the benefits of the exploration and use of space for peaceful purposes, in order to meet the needs of the present generations while preserving **both the Earth and** the outer space environment for future generations.*

***Space sustainability also requires promoting the use and environmental benefits of space data and recognising the need for the launch and in-orbit activities to be carried out in an increasingly responsible and sustainable manner.”***

The new definition is important because it highlights that there is distinct trade-off to be made between the acquisition of space data and the responsible management of space activities. However, looking more closely at the updated definition, it is clear that moving towards a sustainable space economy covers many different aspects. In this regard, the above definition has a striking resemblance to the Brundtland definition of sustainable development [22], which has become which has become synonymous with the concept of triple bottom line (TBL) sustainability [23]. TBL sustainability encompasses environmental, social and economic considerations which are underpinned within the 2030 Agenda for Sustainable Development. However, although traditionally environmental, social, and economic themes are seen to be equal to one another, Giddings, et al. (2002) [24] argues that environmental limits must be respected regardless of social or economic progress. Therefore, when comparing the improved definition against the issues covered by the COPUOS guidelines, the 2030 Agenda for Sustainable Development, the Space Scotland sustainability and net-zero roadmap for the Scottish space sector, the proposal of Giddings, et al. (2002) and general observable trends within the sector, we consider that it covers three broad terms which must be taken into consideration:

1. Sustainability from space: using space as a platform to directly or indirectly address global problems.
2. Sustainability in space: viewing space as a natural resource for preservation, exploitation and exploration.
3. Sustainability for space: protecting the terrestrial environment from the impacts of space activities.

To gain a deeper understanding of each term, it is important to define what issues that they refer to. In this regard, point 1 refers to how the space sector is able to contribute to sustainable development through the provision of space data or more proactive approaches such as energy from space. Here the basic idea is space as a service. As the new space economy develops, it is becoming increasingly more reliant on areas such as EO, navigation and communication for applications such as climate monitoring, urban development, fair trade and equity & justice. Point 2 refers to the sustainable use of space. Here the basic idea is space as a resource which can be preserved for scientific or political reasons, exploited for the benefit of mankind, or explored to help us understand our place in the universe. As such, this covers aspects such as space safety, space environment management, space debris detection & mitigation, in-orbit servicing, in-orbit manufacturing, in-orbit recycling, sustainable use of space resources, re-entry & demise and regulatory & legal aspects. Finally, point 3 refers at the sustainability of the space sector itself based on how it impacts the Earth environment. Potential impact areas cover the entire life cycle of a space mission across raw material extraction, manufacturing, launch, data processing & exploitation, disposal and socio-economic impacts. Both point 2 and point 3 can be combined to form a singular term: sustainability of space. This refers to the management of ecological or other sustainability impacts stemming from space sector activities to both the orbital and Earth environment.

### 3. An Impending Tragedy of the Commons for Space?

Since the beginning of the space race, outer space has generally been considered as a global commons [25]. A global commons refers to resources that are shared and accessible to all, with no single governing country or power [26]. International law identifies four global commons, namely the high seas, the atmosphere, Antarctica and the outer space [27, 28] (even though many states do not view space as such for strategic purposes and national interests, including the USA [29]). With particular reference to space, the global commons principle is reflected in many space laws including the Outer Space Treaty (1967) [30] where Article 1 states that "exploration and use of outer space, including the moon and other celestial bodies, shall be carried out for the benefit and in the interests of all countries... and shall be the province of all mankind'. This is also enshrined within Article 11 of the Moon Agreement (1979) [31] which states that the use of the moon "shall be the province of all mankind and shall be carried out for the benefit and in the interests of all countries, irrespective of their degree of economic or scientific development" and that "the moons and its natural resources are the common heritage of mankind and states may explore and use the moon without discrimination".

Despite this, the landscape of the space sector is vastly different compared to the time when these texts were written. However, from an astropolitics perspective, the big three nations (China, Russia, USA) still dominate and are consistently trying to strategically positioning themselves to be the major global space power [32]. Whilst the long-term vision and aspirations of these players generally do recognise the need for sustainability [33, 34, 35], their actions speak of a different story, with an increasing prospect of the militarisation of space. With this in mind, other states and space blocs are beginning to emerge, with goals of increasing their own autonomy, sovereignty and domestic capacity. Beyond the ambitions of states, New Space has also meant that more states and private companies have been able to access space, led by a significant reduction in launch costs over the last decade [36]. This brings about its own issues, notably concerning the potential (and now likely) scaling up of space sector operations which has the potential to lead to a 'tragedy of the commons' type of scenario playing out in space [37].

The tragedy of the commons is a concept described in economics and ecology whereby common resources (which are not regulated by formal rules) become depleted through individual actions which neglect the well-being of society in the pursuit of personal gain [38]. Whilst space is covered by laws, agreements and guidelines, it is generally considered that many of the leading laws, agreements guidelines have become outdated and are no longer fit for purpose, with many players not seeing themselves as being bound by such agreements anyway. There are five international treaties underpinning space law which are overseen by UN COPUOS. These are the Outer Space Treaty (1967) [30], the Rescue Agreement (1968) [39], the Moon Agreement (1979) [31], the Liability Convention (1972) [40] and the Registration Convention (1976) [41]. However, none of these specifically address the concept of space sustainability due to the novelty of the concept at the time which they were written. To address this gap, a number of international space sustainability guidelines have been published, although these are only voluntary. Such initiatives include the UN COPUOS Long-term Sustainability of Outer Space Activities [9], the ISO 24113:2019 Space Debris Mitigation Requirements [42] and the Inter-Agency Space Debris Coordination Committee (IADC) Space Debris Mitigation Guidelines [14]. The latter or these stipulates that satellites should be removed from orbit within 25 years of end of mission, which is widely considered to be too lax by space sustainability experts in the industry – a common theme amongst the many space sustainability guidelines which exist. The reason for this laxity is that the number and variety of space actors has increased in the decades since the guidelines were developed. Other types of incentives have also been proposed such as the Space Sustainability Rating (SSR) system [43]. The SSR is a composite indicator analogous to the energy rating on consumer electronics and provides a single figure of merit capturing the sustainability of a given satellite mission. Satellite operators can benefit from reduced insurance premiums depending on their score. However, all of these space sustainability guidelines and initiatives mainly focus on preserving the space environment rather than protecting Earth's environment. At present, there does not seem to be a sense of urgency for addressing the environmental impact from the space sector to Earth, as it is not considered to be a strategically important issue.

Furthermore, the legal systems we have in place regarding space are nowhere near as comprehensive as in other spheres such as maritime law. Technological advances have also outpaced space law to a point where space is becoming like the Wild West and no one can be held legally accountable for their actions [32]. This is particularly relevant with New Space as states and private enterprise look at space as a method for commercial or strategic gains. Whilst astropolitics and business issues are beyond the scope of this paper, they do both play a pivotal role with regard to how the industry must evolve in the future to ensure sustainability. In this regard, improving the sustainability of the space sector in orbit requires a complicated, interrelated set of technical, economic, and political challenges to be addressed. This needs to be regulated, without which it is very likely that a 'tragedy of the commons' scenario will occur in space, whereby our inability or failure to manage Earth's orbits as a global common will undermine safety and predictability, leaving space inaccessible to future generations. As such, there is a clear need for an international body such as the UN to maintain a central role in managing outer space affairs and promoting the sustainable use of space. This would require the endorsement of a variety of stakeholders, rather than merely imposing regulations on the space industry without their involvement. Additionally, there is no shortage of agreed-upon space sustainability guidelines at both national and international level, but these are not legally binding and rely on the voluntary goodwill of satellite operators. It may be possible to encourage a limited amount of sustainable behaviour through collaboration and funding opportunities that support the development of sustainable technologies or metrics/targets for space sustainability. However, just as with the major space laws, these incentives have also been perceived by many as being too Westernised for global uptake.

Regardless, it is clear that the legal systems, guidelines and initiatives currently in place are not a long-term solution to space sustainability and could be virtually ineffective at avoiding humanity from falling victim to a tragedy of the commons in outer space if something more comprehensive is not put into place. The lack of oversight with regard to the terrestrial impacts of space activities is also a key ingredient of this potential future, a concept which is encapsulated by the space sustainability paradox.

## 4. The Space Sustainability Paradox

Despite presenting an updated definition of the term 'space sustainability', the Space Scotland definition presents a new problem. The addition of the second paragraph mentions certain activities that might be considered as being either difficult to achieve in tandem or completely inharmonious with one another given their conflicting focus. In particular, the use of space as a service or as a resource will always produce some form of adverse impact on the terrestrial environment given current manufacturing processes and will likely lead to increased levels of orbital congestion and possibly space debris. The more space is used therefore has a distinct threat to push the environment beyond its limits, a prospect which is best demonstrated by Wilson et al. [44] and Miraux et al. [45]. In these studies, it is shown that the sustainability footprint of the space sector could become significant in the near-future, to a point whereby the sector develops into a major contributor to global environmental change. This suggests a distinct and important trade-off between sustainability from space and sustainability of space.

We define this situation as the 'space sustainability paradox', a phrase first coined in November 2021 as part of a two-part lecture seminar at a COP26 side event in Glasgow [46, 47]. Specifically, the term refers to a situation where the increased use of space to support sustainable development leads to heightened adverse impacts to both the space and Earth environment. Thereby the more that the space sector implements space applications and technologies aimed at positively contributing towards sustainable development challenges for the benefit of humankind, the less sustainable the space sector becomes. This point is illustrated in Fig 1, which shows the ongoing imbalance between these two areas as exemplified by an emphasis being placed upon sustainability from space over sustainability of space.

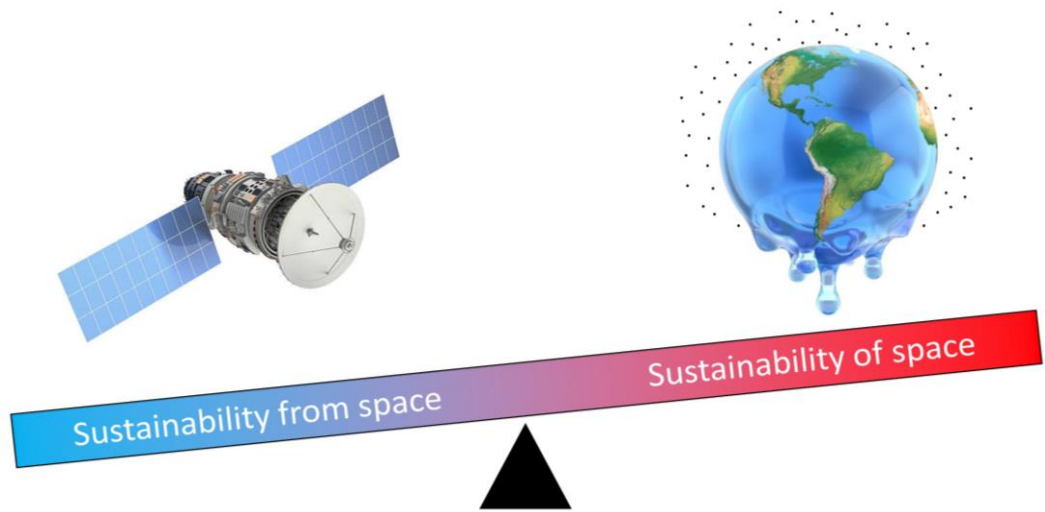

*Fig 1. The space sustainability paradox and the balance between sustainability from space and sustainability of space.*

Key to the space sustainability paradox is the theory that the space sector has a tipping point. Tipping points represent a critical threshold that, when crossed, leads to large and often irreversible changes to the Earth system which drive it into a new state [48]. As outlined in Fig 1, this leads to a fine balance between sustainability from space through an increasing number of satellites and the sustainability of space. Tipping points are also linked with the tragedy of the commons scenario for outer space. In this regard, as indicated in Fig 2, despite future projections indicating that the number of active satellites in orbit will considerably increase in the near future, in the long-term there will be an eventual reduction regardless of whether a business as usual or space environment management approach is applied. This is for one of two reasons: (1) policy comes into place to limit the number of active satellites that can be put into orbit based on carrying capacity and environmental impacts; or (2) a tipping point is reached whereby the Earth and its orbits are placed into a new ecological state, making launching satellites no longer socially, politically or morally acceptable (or even technically possible if Kessler Syndrome is triggered). Such a scenario would therefore suggest that there is a pressing need to define limits by which the growth of space activities should be constrained from an ecological stance. On top this, to avoid a tragedy of the commons scenario from occurring for space, it is essential that space sustainability considerations become a mandatory component of future system definition and design studies and this notion is strategically integrated into space policy.

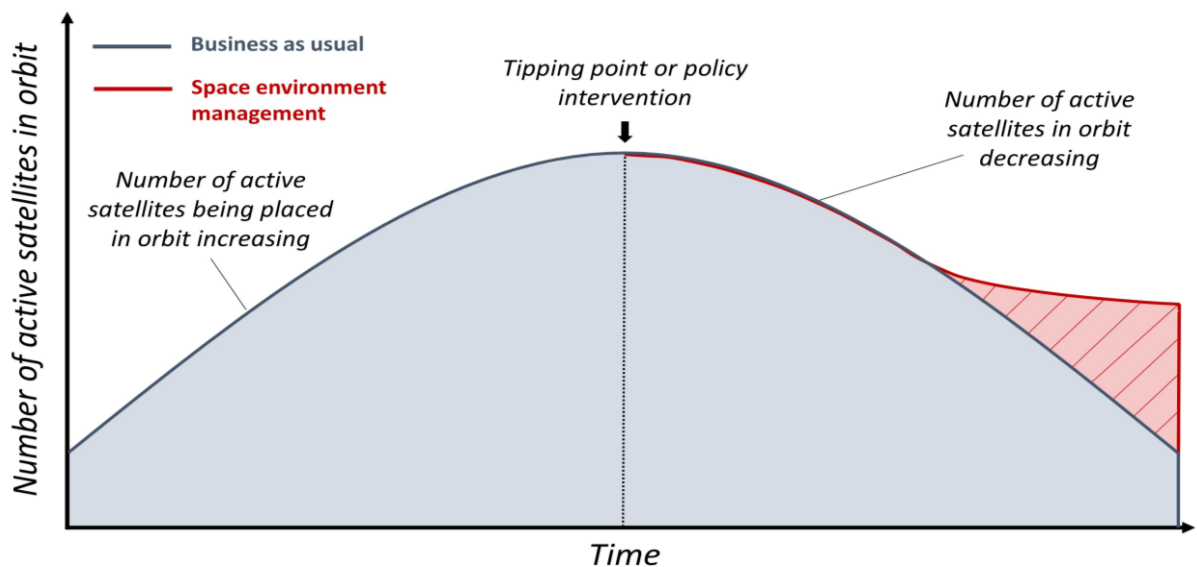

*Fig 2. Illustrative diagram portraying limits of growth in relation to the total number of active satellites in orbit over time.*

However, as will be further discussed in the following section, a key factor to the space sustainability paradox is unintended consequences of missions and policies that target a specific aspect of space sustainability, which often lead to adverse impacts. This kind of situation, therefore, means that sustainability claims made by industrial stakeholders and national agencies should undergo increased scrutiny. In other words, it is not enough to claim sustainability merely through the provision of space data and/or compliance with space debris mitigation guidelines, without considering the wider picture. This is because actions targeting a specific space sustainability aspect without regard to all three terms outlined in Section 2 do not equate to responsible space environment management and in such cases, the entity responsible will have succumb to the space sustainability paradox.

**5. Issues Exacerbating the Problem**

Other sectors have grown and developed without attention to sustainability leading to the current resource and environmental crises. The space sector now has the chance to learn from the other sectors to avoid the same fate and lead by example. However, the space sustainability paradox is a looming threat, given the clear disconnect on the interrelationship between sustainability from space and sustainability of space. This disconnect means that, in many cases, even our space sustainability activities can exacerbate this problem. Therefore, in order to avoid falling victim to the space sustainability paradox, a radical rethink on how to address space sustainability challenges is required. To start, it is important that we consider the key issues that intensify the space sustainability paradox from the viewpoint of space being both a source and receptacle of potential anthropogenic pollution. This has three spheres of influence: the Earth environment, Earth orbit or beyond Earth orbit (and into deep space). However, with respect to the space sustainability paradox, only the key issues associated with the interrelationship between the Earth environment and orbital environment will be discussed in the remainder of this section.

Firstly, the space sustainability paradox is exacerbated by a lack of meaningful action on addressing the environmental footprint of the space sector. Having net-zero goals, producing ESG reports (particularly in cases where the benefits of space data is listed without an indication into the environmental footprint of these activities) and holding space sustainability events is nothing more than meaningless high-level jargon without discernible action resulting in tangible and/or quantifiable results. As a result, the term 'space sustainability' has increasingly found its way into the mainstream media as the new buzzword of the space industry, despite many stakeholder declarations or actions lacking any real substance or depth. For example, phasing out hydrazine for high performance green propellants (HPGPs) does not equate to sustainability. The use of the term 'green' in this sense does not mean something less harmful to the environment, but merely that the propellant is non-toxic – a misalignment with the terminology used in other industries. In fact, a study conducted by the University of Strathclyde which investigated the environmental impact of replacing traditional propellants with HPGPs found that LMP-103S (a HPGP) performed worse than hydrazine across 13 out of 16 environmental impact categories [49], a result which was mirrored by an ESA study conducted by Thales Alenia Space [50]. Another example centres around the verbiage that launcher reusability is sustainable. Many of these claims are unfounded given that a life cycle assessment (LCA) study has never been completed. In reality, reusability does not inherently mean green. Whilst reusability reduces the dependency on virgin materials and prevents rocket stages coming to rest on the ocean floor, more propellant is required, heightened levels of transportation is necessary (as part of recovery operations) and there is an increase in the use of energy for stage refurbishment. Hence, there is a clear trade-off which needs to be made as to whether this is truly a green option considering a range of environmental media, in a similar manner to the study of Dominguez Calabuig, et al. (2022) [51]. Therefore, we propose that a sustainable solution must be one which is quantifiable, with a proven reduction in impact beyond that of merely $CO_2$ and/or GHGs (described as carbon tunnel vision). This also includes caution around the potential future use of carbon offsetting - a practice which should not be used as a licence to pollute or undermine reduction measures. Therefore, to claim sustainability, space sustainability policies and actions should present results which are verifiable and transparent, based on a robust and reliable methodology. In this regard, making unfounded high-level statements without scientific proof or quantification cannot be described as a sustainable solution and would verge on greenwashing. Despite this, in the long-term, ensuring a better environmental footprint will probably mean lessening the number of future space missions. This is because whilst space sustainability tools are useful for addressing environmental impacts at an efficiency level, they do little to address the ongoing growth in space activities. This issue must be addressed at a political level, taking into account tipping points and perhaps limiting the amount of space missions that can be launched by strategic importance. This includes avoiding launching spacecraft whose goals could realistically be fulfilled by existing/proposed space missions or whose addition would add little value to the status quo.

In terms of the orbital space environment, the increasing number of actors and objects in space have resulted in an increased risk of satellite collisions, as well as making coordination and space situational awareness more challenging. This is particularly problematic as the space debris population is not well-characterised and smaller pieces of debris cannot be tracked accurately, with space debris mitigation techniques being costly and technically challenging. However, similar to Earth, we can also view Low Earth Orbit (LEO) as self-regulating – that being that all objects will eventually fall back to Earth, albeit very gradually [52]. In reality, from an environmental perspective, the most sustainable option would actually be to end space operations altogether and allow space debris to re-enter naturally, thereby allowing the space environment to cleanse itself. However, this is not seen to be economically advantageous or socially/politically acceptable given that the benefits stemming from space has become an integral part of modern living. The issue is that since spaceflight began, we have been contributing to the problem at a faster rate than our orbits can self-regulate, making space operations more dangerous. Therefore, we must learn to use space in a more sustainable way, respecting the ecological thresholds of both Earth and our orbital environment. Be that as it may, the current situation is that the space debris problem coupled with the expansion of space-based services is limiting the availability of orbital slots. This is compounded by an absence of comprehensive international regulations and coordination frameworks for space activities, which adds complexity to the management and governance of orbital space. Addressing these challenges requires concerted efforts in debris mitigation, space traffic management and international cooperation. But in reality, rapid advancements in space technology and the commercialisation of space activities often leads to the prioritisation of short-term objectives over long-term sustainability considerations.

However, we are likely to make mistakes if we only look at these spheres individually, as the solutions we propose to address challenges in a specific sphere may have unintended consequences on the other spheres. One example of this is the general compliance of space missions with the IADC Space Debris Mitigation Guidelines [14]. Such space missions may be de-orbited from LEO by directly re-entering the spacecraft into Earth's atmosphere, thereby removing it from orbit in full compliance with the guidelines. However, this action pollutes the upper atmosphere with metallic compounds. On this, a team at the Massachusetts Institute of Technology (MIT) conducting research into measuring particles in the stratosphere found a "menagerie of metal compounds", including lead, zinc and copper, that may point to a further human effect on cloud formation [53]. Whilst likely sources of these compounds come from industrial activities such as smelting and open-pit burning of electronics, Dr Martin Ross of the Aerospace Corporation suspects that some of this comes from spacecraft re-entries [54]. He hypothesises that re-entry smoke particles "have unusual cloud condensation behaviour that could make them more important for climate or ozone". Whilst this is not so much of a problem now given that far more meteoroids enter Earth's atmosphere than spacecraft, this could soon change with the prospect of mega-constellations, space tourism, space colonisation, space solar power, etc. This would suggest that existing policies and remediation actions to dispose of space debris are not a long-term sustainable solution and new technologies and guidance are required to reuse and recycle our assets in space in a similar fashion in which recycling garbage is a viable solution on Earth. But more broadly, it indicates that just because something has sustainable motives does not mean that it is sustainable. We must consider the entire outcome of a proposed solution before jumping on the sustainability bandwagon.

6. **Rethinking Space Sustainability**

Undoubtedly, given the severity of the space sustainability paradox, there is an increasing pressure to resolve the tension between the growth of the space sector and its environmental and socio-economic impact. The private sector has grown at an accelerated rate in the last decade with the New Space movement generating new commercial services at a fast pace. This rapid growth has brought with it a significant economic return but with a non-negligible and intensifying impact on the Earth and space environment. This negative impact creates a conflict between the benefits coming from new space applications and the sustainable growth of the space sector. It is therefore paramount that this is resolved in a comprehensive and transparent manner to prevent exhausting the two primary resources that space can provide: our most important life support system (Earth) and space itself.

To do this, and since space is a domain that acts as a platform for addressing the social foundation, it is important that its limits are as respected as a planetary boundary. Planetary boundaries were first proposed by Rockstrom, et al. (2009) [55] to highlight anthropogenic perturbations of the Earth system in relation to safe operating thresholds/tipping points. The first-time planetary boundaries were linked to space activities from a sustainability of space perspective was in Wilson (2019) [49], but it is now looking increasingly likely that this approach will be used as a measurement/benchmark of space-related impact in the future. However, with respect to the sustainability paradox, there is a need to go beyond this view to also address the social foundation for which space activities support. In this regard, the doughnut of social and planetary boundaries model proposed by Raworth (2017) [56] presents a viable approach. As shown by Raworth (2017), the Earth system has an ecological ceiling which should be maintained to ensure that humanity does not collectively overshoot the nine planetary boundaries that protect Earth's life-supporting systems. Similarly, the model also stresses the importance of addressing humanity's social foundation through twelve basic human needs for avoiding critical deprivation, thereby ensuring that no one is left falling short on life's essentials. In essence, a certain amount of environmental impact is unavoidable if we want to meet the social needs of people, but between the two sets of boundaries lies a doughnut-shaped space that we should strive to conserve – a space which is both ecologically safe, socially just and where humanity can thrive. However, in both the planetary boundaries model and doughnut of social and planetary boundaries model, space is missed out. This is a major omission, given that space sustainability needs to fall into this safe and just space through the provision of data to support the social foundation without breaching the ecological ceiling. Therefore, in line with the 'SDG 18: Space for All' concept [57], it would be beneficial to consider orbital space depletion as the tenth planetary boundary and use this as an instrument for evaluating the complete ramifications stemming from space sustainability policies and actions, as exemplified in Fig 3.

The inclusion of orbital space depletion in Fig 3 is important because, if left unchecked, the orbital space environment could reach a tipping point or critical threshold whereby the orbital population becomes so saturated that the negative impacts of future proposed spacecraft begin to outstrip the benefits of the intended application. In this regard, orbital space depletion is linked with orbital carrying capacity. Orbital carrying capacity can be defined as the maximum amount space objects that can sustained by a particular orbit, given the amount of orbital space available [58, 59]. Orbital space depletion therefore refers to the increasing occupational volume of a given orbit considering the number of operational satellites and the local space debris population size [60]. Therefore, with a decrease in orbital resource availability there is an enhanced risk of collision and the propagation of new clouds of debris. Although a specific parameter has not been formally defined for orbital space depletion as an ecological threshold yet, the basic idea is that someday we could exceed orbital capacity whereby certain bands of orbit become so crowded with satellites and space debris that they become difficult to navigate. This parameter is critical for defining the maximum number of active satellites that can be sustained in space long-term, so that the trend line in Fig 2 plateaus under the space environment management scenario rather than recedes under the business-as-usual scenario. More research is needed to define the critical threshold and/or tipping point for the space resource depletion planetary boundary, as well as the contribution of the space sector towards the social foundation and each of the nine other planetary boundaries.

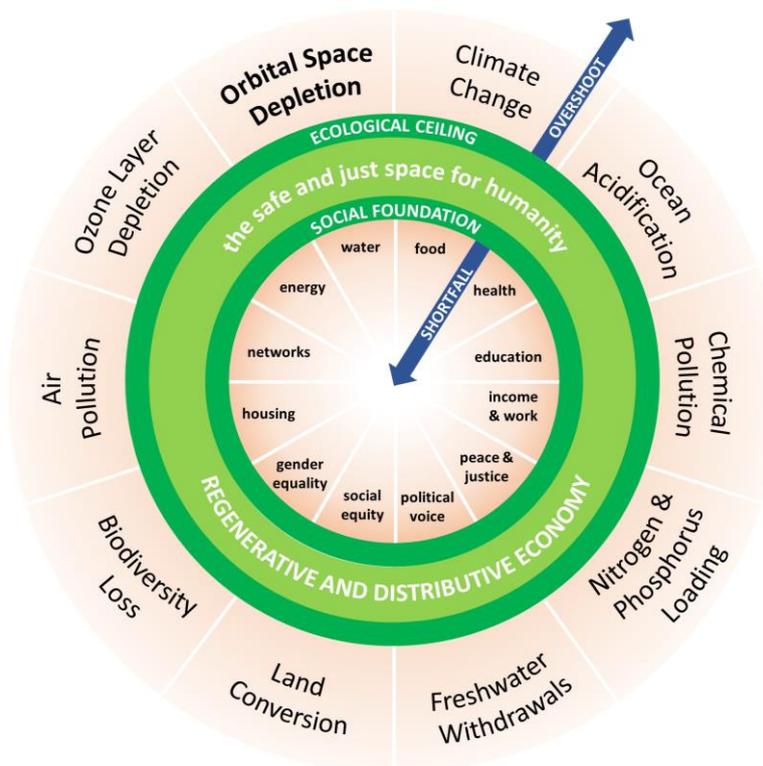

*Fig 3. The doughnut of social and planetary boundaries depicting orbital space depletion as the tenth planetary boundary (adapted from Raworth, 2017 [56]).*

However, based on Fig 3, it is critical that the interrelated dynamics of a space mission between each planetary boundary are properly respected and evaluated when assessing the sustainability profile of a space mission. For instance, in the space debris mitigation example mentioned earlier in this section, whilst re-entering spacecraft at the end of life will lessen pressure on the orbital space depletion boundary, as stated it will have a profound impact on atmosphere, increasing pressures on other planetary boundaries such as climate change and ozone layer depletion. But additionally, besides this atmospheric pollution, the resource value will be lost forever, increasing pressure on many other planetary boundaries and the social foundation indirectly through the requirement of using virgin materials in the build of future space missions to address other social issues. All of this is not to understate the importance of space data but does imply that consideration should be given to the fact that the act of launching and using satellites to aid the social foundation may have the opposite effect on the ecological boundaries. In other words, this could just be replacing one problem with another. We therefore have to ensure that we stay within the 'safe and just space for humanity' doughnut, defining the role of the space sector towards both the social foundation and planetary boundaries.

Therefore, how to counterbalance sustainability from space and sustainability of space becomes important if a tragedy of the commons scenario is to be avoided within the space sector. To achieve this, it is clear that new avenues will need to be created to facilitate multi-stakeholder dialogue and international cooperation to promote and strengthen stability, security and regulation arounds space sustainability issues. These discussions should focus on creating an international framework which recognises space as a global commons with tipping points, that needs to be effectively managed. Setting goals is a good place to start, and based on Fig 3, this could include targeting net-zero and a circular space economy by 2050 to fall in line with other sectors. In this context net-zero should encapsulate three different zeros:

- Zero debris proliferation to maintain the necessary orbit carrying capacity;
- Zero adverse impacts to the Earth environment to avoid breaching planetary boundaries; and
- Zero unsustainable use of resources to enable both above points to be achieved.

This vision of a circular space economy signifies the use and re-use of space resources, including in orbit servicing, repair, manufacturing and recycling. It also ensures the responsible management of space debris, resources, environmental impact mitigation and equitable access to space-based opportunities. In this sense, it should firstly prioritise the minimisation and mitigation of space debris, aiming for zero proliferation through stringent debris mitigation measures, active debris removal initiatives and responsible end-of-life practices. Secondly, the framework should strive for net-zero adverse impacts on the Earth environment by minimising emissions and other negative environmental consequences associated with space activities. This entails adopting eco-design principles, promoting transparency/openness in reporting and implementing sustainable operational practices. Lastly, the framework should ensure zero unsustainable use of resources by fostering a circular economy approach, promoting in-orbit servicing, repair, manufacturing and recycling by means of efficient resource utilisation. Through the integration of modular design principles, the disassembly and repurposing of satellites and space structures become achievable objectives, as exemplified through the CORES project [61]. Such a scenario could transform previously undesirable or problematic materials into valuable feedstock for future manufacturing processes, paving the way for the achievement of the net-zero and circular space economy vision via the development of new and effective space sustainability business models.

Through the three inter-connected net-zero principles, the space industry can begin to chart a sustainable path forward, safeguarding the space environment, minimising its impact on Earth, and optimising resource utilisation for long-term viability and prosperity of the industry. Such principles are highly aligned to the Spaceship Earth concept which is a worldview that Earth is a self-contained system, dependent on its own vulnerable supply of natural resources, and requires everyone to act as a harmonious crew working toward the greater good [62]. Whilst the future of space exploration may ultimately change this perspective (from closed-loop to open-loop), the central theme of needing to work together to maintain the safety and sustainability of our environment remains a constant, particularly given that our environment can no longer be considered only as Earth, but our whole solar system [63]. To do this, it is important that all future space policies or actions incorporate these principles and are examined through the adapted version of the doughnut of social and planetary boundaries presented in Fig 3 before they are approved. However, given that the sustainable use of resources in space implies that less stress is placed upon both the orbital and Earth environment, creating remanufacturing facilities in space could be considered as a crucial underpinning element and starting point of this proposed framework in the pursuit of space sustainability to avoid a tragedy of the commons in space.

A comprehensive space sustainability framework of this kind is integral for paving the way towards a circular space economy that operates within the safe and just boundaries of our planet, while unlocking the potential of space for all. This is important to ensure that the benefits of space exploration and utilisation are equitably distributed, promoting inclusivity and addressing social and economic inequalities. The responsibility of this will require international oversight, so the UN would be a good mechanism for producing this framework and igniting the pathway for this technical and regulatory transformation. Such a framework could build upon the Guidelines for the Long-term Sustainability of Outer Space Activities or look into the feasibility of recognising space as a global commons (or 18th SDG) in the post-2030 Agenda for Sustainable Development. Collectively, these actions would put us on a path towards guaranteeing the sustainable use of space, the sustainable use of space resources and a sustainable impact to the Earth and space environment.

## 7. Conclusion

Space sustainability is a growing concept with a limited and often misconstrued focus. As demonstrated within this paper, this is leading to the emergence of a paradoxical situation whereby the use of space to support sustainability is becoming unsustainable. We argue that not enough is currently being done to tackle this problem and that a momentous paradigm shift may be required in order to prevent negative trends from becoming the norm. To ease tension between sustainability from space and sustainability of space, space activities and policies must respect the interrelated nature of space sustainability issues. One way to do this is to consider space sustainability through the doughnut of social and planetary boundaries framework, with orbital space depletion integrated as the tenth planetary boundary. Such an approach may help to distinguish the right balance between social benefits from space and the space sector's growing contribution towards ecological thresholds. Only when this is achieved can the sector truly begin to achieve sustainability and ensure the continued use of outer space by future generations.

**References**

[1] A. R. Wilson, S. M. Serrano, K. J. Baker, H. B. Oqab, G. B. Dietrich, M. Vasile, T. Soares and L. Innocenti, "From Life Cycle Assessment of Space Systems to Environmental Communication and Reporting," in *72nd International Astronautical Congress*, Dubai, United Arab Emirates, 2021.

[2] United Nations General Assembly, "Transforming our world: the 2030 Agenda for Sustainable Development," Resolution adopted by the General Assembly on 25 September 2015 (A/RES/70/1), New York, USA, 2015.

[3] H. Du, "Space for the Sustainable Development Goals (Space4SDGs)," in *Eighth Meeting of the Inter-Agency and Expert Group on Sustainable Development Goal Indicator (IAEG-SDGs)*, Stockholm, Sweden, 2018.

[4] The Guardian, "Billionaire space cowboyscould become heroes by focusing on the climate crisis," 25 July 2021. [Online]. Available: https://www.theguardian.com/business/2021/jul/25/billionaire-space-cowboys-could-become-heroes-by-focusing-on-the-climate-crisis. [Accessed 23 May 2023].

[5] P. Sutter, "Megaconstellations could destroy astronomy and there's no easy fix," Space.com, 23 May 2023. [Online]. Available: https://www.space.com/megaconstellations-could-destroy-astronomy-no-easy-fix. [Accessed 06 October 2021].

[6] J. O'Callaghan, "Satellite Constellations Could Harm the Environment, New Watchdog Report Says," Scientific American, 24 November 2022. [Online]. Available: https://www.scientificamerican.com/article/satellite-constellations-could-harm-the-environment-new-watchdog-report-says/. [Accessed 23 May 2023].

[7] A. Hirschfeld, "What Is the Environmental Impact of the Billionaire Space Race? Experts Weigh In," The Observer, 28 July 2021. [Online]. Available: https://observer.com/2021/07/environmental-impact-billionaire-space-race-experts-bezos-musk-branson/. [Accessed 23 May 2023].

[8] B. Detsis and E. Detsis, "The benefits brought by space—General public versus space agencies perspectives," *Acta Astronautica,* vol. 88, pp. 129-137, 2013.

[9] United Nations Committee on the Peaceful Uses of Outer Space, "Guidelines for the long-term sustainability of outer space activities," Fifty-fifth Scientific and Technical Subcommittee (A/AC.105/2018/CRP.20), Vienna, Austria, 2018.

[10] C. D. Johnson and L. G. Croy, "Handbook for New Actors in Space," Secure World Foundation, 2017.

[11] Secure World Foundation, "Space Susainability - A Practical Guide," 2014.

[12] European Space Agency, ""ESA and the Sustainable Development Goals," [Online]. Available: https://www.esa.int/Enabling_Support/Preparing_for_the_Future/Space_for_Earth/ESA_and_the_Sustainable_Development_Goals . [Accessed 23 May 2023].

[13] A. Baumgart, E. I. Vlachopoulou, J. D. Rio Vera and S. Di Pippo, "Space for the Sustainable Development Goals: mapping the contributions of space-based projects and technologies to the achievement of the 2030 Agenda for Sustainable Development," *Sustainable Earth Reviews,* vol. 4, no. 6, 2021.

[14] Inter-Agency Space Debris Coordination Committee, *Space Debris Mitigation Guidelines,* Forty-Fourth Session of the Scientific and Technical Subcommittee (A/AC.105/C.1/L.260): Committee on the Peaceful Uses of Outer Space, 2007.

[15] D. McKnight, *University space research: Evolutionary versus revolutionary technology imprevatives with practical observations about orbital debris,* Public guest lecture: University of Strathclyde, 2016.

[16] D. J. Kessler, N. L. Johnson, J. Liou and M. Matney, "The Kessler Syndrome: Implications to Futue Space Operations," *Advances in the Astronautical Science,* vol. 137, no. 8, 2010.

[17] T. Maury, P. Loubet, S. M. Serrano, A. Gallice and G. Sonnemann, "Application of environmental life cycle assessment (LCA) within the space sector: A state of the art," *Acta Astronautica,* vol. 170, pp. 122-135, 2020.

[18] A. Lawrence, M. L. Rawls, M. Jah, A. Boley , F. Di Vruno, S. Garrington, M. Kramer, S. Lawler, J. Lowenthal, J. McDowell and M. McCaughrean, "The case for space environmentalism," *Nature Astronomy,* vol. 6, no. 4, pp. 428-435, 2022.

[19] M. Palmroth, J. Tapio, A. Soucek, A. Perrels, M. Jah, M. Lönnqvist, M. Nikulainen, V. Piaulokaite, T. Seppälä and J. Virtanen, "Toward Sustainable Use of Space: Economic, Technological, and Legal Perspectives," *Space Policy,* vol. 57, no. 101428, 2021.

[20] L. Miraux, "Environmental limits to the space sector's growth," *Science of The Total Environment,* vol. 806, no. 4, p. 150862, 2022.

[21] Space Scotland, Scottish Enterprise, Astrogency and Optimat, " Space Sustainability: A Roadmap for Scotland," 2022.

[22] World Commission on Environment and Development, "Our common future," Oxford University Press, Oxford, 1987.

[23] J. Elkington, "Cannibals with forks: the triple bottom line of twenty-first century," New Society Publishers, 1997.

[24] B. Giddings, B. Hopwood and G. O'brien, "Environment, Economy and Society: Fitting Them Together into Sustainable Development," *Sustainable Development,* vol. 10, pp. 187-196, 2002.

[25] C. Delaune, "Outer Space as a Global Common: Toward Tragedy or Governance," *Journal of Student Research,* vol. 11, no. 4, 2022.

[26] J. Vogler, "Global Commons Revisited," *Global Policy,* vol. 3, no. 1, pp. 61-71, 2012.

[27] I. Sarenmalm, "Sustainable Development in International Law and the protection of the Global Commons," Master thesis in Sustainable Development, Uppsala University, 2017.

[28] D. Garcia, "Global commons law: norms to safegaurd the planet and humanity's heritage," *International Relations,* vol. 35, no. 3, pp. 422-445, 2021.

[29] Exec. Order No. 13914, 85 Fed. Reg. 20,381 (Apr. 10, 2020)..

[30] Treaty on Principles Governing the Activities of States in the Exploration and Use of Outer Space, including the Moon and other Celestial Bodies, adopted on 19 December1966, opened for signature on 27 January 1967, entered into force on 10 October 1967, 1967.

[31] Agreement Governing the Activities of States on the Moon and Other Celestial Bodies, UN Doc. A/34/664 , 1979.

[32] T. Marshall, The Future of Geography: How Power and Politics in Space will Change Our World, London, UK: Elliott and Thompson Limited, 2023.

[33] The State Council Information Office of the People's Republic of China, "China's Space Program: A 2021 Perspective," China National Space Administration, 28 January 2022. [Online]. Available: http://www.cnsa.gov.cn/english/n6465645/n6465648/c6813088/content.html. [Accessed 23 May 2023].

[34] National Aeronautics and Space Administration, "NASA Strategic Plan 2022," 28 March 2022. [Online]. Available: https://www.nasa.gov/sites/default/files/atoms/files/2022_nasa_strategic_plan.pdf. [Accessed 23 May 2023].

[35] F. Vidal, "Russia's Space Policy: The Path of Decline?," The French Institute of International Relations, Paris, France, 2021.

[36] V. L. Shammas and T. B. Holen, "One giant leap for capitalistkind: private enterprise in outer space," *Palgrave Communications,* vol. 5, no. 10, 2019.

[37] P. Wang, "Tragedy of the Commons in Outer Space - The Case of Space Debris," in *64th International Astronautical Congress*, Beijing, China, 2013.

[38] G. Hardin, "The tragedy of the commons," *Science,* vol. 162, pp. 1243-1248, 1968.

[39] Agreement on the Rescue of Astronauts, the Return of Astronauts and the Return of Objects Launched into Outer Space, adopted on 19 December 1967, opened for signature on 22 April 1968, entered into force on 3 December 1968, 1968.

[40] Convention on Liability for Damage Caused by Objects Launched into Outer Space, adopted on 29 November 1971, opened for signature on 29 March 1972, entered into force on 1 September 1972, 1972.

[41] Convention on Registration of Objects Launched into Outer Space, adopted on 12 November 1974, opened for signature on 14 January 1975, entered into force on 15 September 1976, 1976.

[42] ISO 24113:2023 , *Space Systems - Space debris mitigation requirements,* Geneva, Switzerland, 2023.

[43] A. Saada, E. David, F. Micco, J.-P. Kneib, M. Udriot, D. Wood, M. Rathnasabapathy, D. Weber, F. Letizia, S. Lemmens, M. Jah, S. Potter, N. Khlystov, M. Lifson, K. Acuff, R. Steindl and M. Slavin, "The Space Sustainability Rating: An operational process incentivizing operators to implement sustainable design and operation practices," in *73rd International Astronautical Congress*, Paris, France, 2022.

[44] A. R. Wilson, M. Vasile, C. A. Maddock and K. J. Baker, "Ecospheric life cycle impacts of annual global space activities," *Science of The Total Environment,* vol. 834, no. 155305, 2022.

[45] L. Miraux, A. R. Wilson and G. J. Dominquez Calabuig, "Environmental sustainability of future proposed space activities," *Acta Astronautica,* 2022.

[46] M. Vasile and A. R. Wilson, *The Space Sustainability Paradox: From Sustainability from Space to the Sustainability of Space,* Glasgow: Sustainable Glasgow Landing Hub, 2021.

[47] A. R. Wilson and M. Vasile, *What is Space Sustainability And How Can It Be Achieved?,* Glasgow: Sustainable Glasgow Landing Hub, 2021.

[48] Intergovernmental Panel on Climate Change, "Climate Change 2021: The Physical Science Basis. Contribution of Working Group I to the Sixth Assessment Report of the Intergovernmental Panel on Climate Change," Cambridge University Press, Cambridge, United Kingdom, 2021.

[49] A. R. Wilson, "Advanced Methods of Life Cycle Assessment for Space Systems," PhD thesis, University of Strathclyde, 2019.

[50] N. Thiry, P. A. Duvernois, F. Colin and A. Chanoine, "GreenSat: Final Report," European Space Agency Report, Noordwijk, Netherlands, 2019.

[51] G. J. Dominguez Calabuig, L. Miraux, A. R. Wilson, A. Pasini and A. Sarritzu, "Eco-design of future reusable launchers: insight into their life cycle and atmospheric impact," in *9th European Conference for Aeronautics and Space Sciences*, Lille, France, 2022.

[52] B. Weeden, "Overview of the legal and policy challenges of orbital debris removal," *Space Policy,* vol. 27, no. 1, pp. 38-43, 2011.

[53] D. J. Cziczo, K. D. Froyd, C. Hoose, E. J. Jensen, M. Diao, M. A. Zondlo, J. B. Smith, C. H. Twohy and D. M. Murphy, "Clarifying the Dominant Sources and mechanisms of cirrus cloud formation," *Science,* vol. 340, no. 6138, pp. 1320-1324, 2013.

[54] M. N. Ross, *Personal communication,* 15 February 2018.

[55] J. Rockström, W. Steffen, K. Noone, Å. Persson, F. S. Chapin III, E. Lambin, T. M. Lenton, M. Scheffer, C. Folke, H. J. Schellnhuber, B. Nykvist, C. A. de Wit, T. Hughes, S. van der Leeuw, H. Rodhe, S. Sörlin, P. K. Snyder, R. Costanza, U. Svedin, M. Falkenmark, L. Karlberg, R. W. Corell, V. J. Fabry, J. Hansen, B. Walker, D. Liverman, K. Richardson, P. Crutzen and J. Foley, "Planetary Boundaries: Exploring the Safe Operating Space for Humanity," *Ecology and Society,* vol. 14(2), no. 32, 2009.

[56] K. Raworth, Doughnut Economics: Seven Ways to Think Like a 21st-Century Economist, Random House Business, 2017.

[57] C. Moenter, C. Illgner, L. Selles and L. Ike, "SDG 18 - Space for All," 2023. [Online]. Available: https://sdgspace.wordpress.com/. [Accessed 2023 May 23].

[58] S. Rouillon, *A physico-economic model of space debris management,* Cahiers du GREThA, n°2019-10: University of Bordeaux, 2019.

[59] F. Letizia, S. Lemmens and H. Krag, "Environment capacity as an early mission design driver," *Acta Astronautica,* vol. 173, pp. 320-332, 2020.

[60] T. Maury, P. Loubet, J. Ouziel, M. Saint-Amand, L. Dariol and G. Sonnemann, "Towards the integration of orbital space use in Life Cycle Impact Assessment," *Science of The Total Environment,* vol. 595, pp. 642-650, 2017.

[61] M. A. Laino, A. R. Wilson, M. Vasile, R. Paoli and C. Gales, "A Transport Network for In-Orbit Recycling Exploiting Natural Dynamics," *Acta Astronautica,* pp. Pre-print, 2023.

[62] H. George, George, Henry. Progress and Poverty: An Inquiry Into the Cause of Industrial Depressions, and of Increase of Want with Increase of Wealth, Garden City, New York, USA : Book IV, Chapter 2, Doubleday, Page & Co, 1879.

[63] A. R. Wilson, "An Exploration into the Viability of Space-Based Technology to Combat Climate Change," BSc (Hons) dissertation, Glasgow Caledonian University, 2015.
18